\documentclass{aastex}
\usepackage{emulateapj5}
\usepackage{graphics}
\usepackage{psfig}

\def\R{{\sl ROSAT}}

\def\C{{\sl Chandra}}

\def\N{{\sl XMM-Newton}}

\newcommand\hi{H{\small I}}

\begin{document}

%\submitted{version 5 -- April 24 2001}
\title{\bf Chandra Detection of a Hot Gaseous Corona Around the Edge-on Galaxy 
NGC~4631}

\author{Q. Daniel Wang\altaffilmark{1}, 
Stefan Immler\altaffilmark{1},
Rene Walterbos\altaffilmark{2}, 
James T. Lauroesch\altaffilmark{3} \& 
Dieter Breitschwerdt\altaffilmark{4}}
\affil{$^1$Astronomy Department, University of Massachusetts, Amherst, MA 
01003}
\affil{$^2$Astronomy Department, New Mexico State University, Las Cruces, NM 
88003}
\affil{$^3$Department of Physics and Astronomy, Northwestern University, 
Evanston, IL 60208}
\affil{$^4$Max-Planck-Institut f\"ur extraterrestrische Physik, 85748 
Garching, Germany}

\shorttitle{\C\ Detection of a Hot Gaseous Corona Around the Edge-on Galaxy 
NGC~4631}
\shortauthors{Wang et al.}

\begin{abstract}
We present a \C\ X-ray observation that shows, unambiguously for the
first time, the presence of a giant diffuse X-ray-emitting corona around 
the edge-on disk galaxy NGC 4631. This corona, with a characteristic thermal 
temperature of $2$--$7\times10^6$~K, extends as far as 8~kpc away from the 
galactic plane. The X-ray morphology resembles the radio halo of the galaxy, 
indicating a close connection between outflows of hot gas, cosmic rays, and 
magnetic field from the galactic disk. Enhanced diffuse X-ray emission is 
apparently enclosed by numerous H$\alpha$-emitting loops blistered out from 
the central disk of the galaxy, as is evident in a comparison with our deep
{\sl Hubble Space Telescope} imaging. 
\end{abstract}


\section{Introduction}
Though postulated more than four decades ago (Spitzer 1956), the existence of 
a hot ($T\sim10^6$~K) extended corona around our Milky Way galaxy is still
subject to debate. What limits our ability to understanding the corona
around the Milky Way is our location in its midst. Given the expected
characteristic temperature of $\sim10^6$ degrees for this gas, soft X-ray
emission is the primary tracer for the Galactic corona.
However, the soft X-ray sky observed from our location within the Galaxy
is very confusing due to the multiple origins of X-ray-emitting hot gas.
Any extended Galactic corona, if present, is embedded between a nonuniform
foreground, produced by the Local Hot Bubble around the Sun and various
contributions from other nearby interstellar gas clouds, and an extragalactic
soft X-ray background of an uncertain spectrum.  Disentangling these
components along a typical line of sight is extremely difficult, if not
impossible. 

The most direct way to study extended galactic coronae is to observe nearby
edge-on disk galaxies. \R\ observations of such galaxies have demonstrated
that extra-planar X-ray emission is present in relatively normal spirals
(NGC 891, Bregman \& Houck 1997; NGC 4631, Wang  et al. 1995). 
Such ``normal'' edge-on galaxies, in which star formation is not dominated by 
galactic nuclear regions, but extend over a large fraction of the galactic 
disk,
are ideal for investigating general properties of galactic halos and their
connections to galactic disks. The limited sensitivity and spatial resolution
of \R, however, did not allow for an unambiguous separation of diffuse emission
from discrete sources, let alone a detailed study of morphology, ionization, 
and dynamics of hot gas.

In this {\sl Letter} we report preliminary results from observations of
the SBc/d type galaxy NGC 4631 with the \C\ X-ray observatory and the
{\sl Hubble Space Telescope}. This galaxy is ideal for studying soft
X-ray emission from its galactic corona because of the very low Galactic
foreground absorption ($N_{\rm HI} = 1.2\times10^{20}~{\rm cm}^{-2}$) and
its edge-on orientation with respect to the viewing direction. Active
star formation is apparent throughout much of the NGC 4631 disk,
apparently triggered by interactions with neighboring galaxies.  Evidence
for this interaction includes the presence of \hi\ tidal tails
(Weliachew, Sancisi, \& Guelin 1978).  Such active star formation is presumably
a necessary condition for the heating of coronal gas. 

\section{Chandra Observation and Data Calibration}

Our \C\ observation was carried out with the Advanced CCD Imaging Spectrometer 
chip S3 (ACIS-S3) at the focal plane of the telescope mirror. The spatial
resolution of the observation before any smoothing is $\sim1''$, corresponding
to $\sim37$~pc at the 7.6 Mpc distance of the galaxy. This superb resolution,
together with the high sensitivity of the instrument and the long exposure
($\sim60$~ks), allows us to remove point-like sources and to directly
probe the corona and its relation to other galactic components.  We calibrated
the data using primarily the software package {\sc ciao}
v2.1 with the latest calibration products ({\sc caldb} 
version 2.3). We selected data only in the energy range of 0.3--7~keV.
Particle background is greatly enhanced at higher energies, while the
calibration of both the gain (uncertainties up to $\sim30\%$) and quantum
efficiency (up to $\sim10\%$) of the detector are unreliable at lower energies.
We excluded bad pixels, rows and node boundaries, and screened the data 
for high and
low background times by clipping observation periods with an offset exceeding
$\pm3\sigma$ from the mean quiescent count rate ($1.16 {\rm~cts~s^{-1}}$).
This filtering led to the exclusion of 8\% of the total counts and 6\% of the
total exposure time, resulting in a cleaned exposure time of 55.1~ks.

We further conducted both background subtraction and exposure correction
of the data.  We use deep (138~ks) blank-field background datasets with
similar foreground absorption and identical focal plane temperature as our
NGC 4631 observation to determine the position depended background rates.
The same data filtering criteria 
were applied. We created bad pixel and node-boundary
removed exposure maps in various energy bands. For the spectral analysis of
extended X-ray emission, we constructed weighted instrument 
effective area and spectral response matrices, using a low-resolution 
(bin size $= 4''$), 0.3--2.5~keV band image of diffuse emission in 
detector coordinates. The weighting corrects for variations (up to $\sim20\%$) 
in the instrument sensitivity  across the detector (e.g., effective
area, gain, and quantum efficiency). 
It is important to note that the currently available instrument response 
matrices at low energies ($\lesssim 1$~keV) employ various approximations,
which may not be appropriate and may lead to underestimation of line 
strength. 
Since NGC 4631 fills nearly the entire 
ACIS-S3 field-of-view, we extracted the background from the same detector 
regions in the empty-field background observations described 
above as the source regions in the NGC 4631 observations.

\section{Analysis and Results}

\C\ images in the soft and hard X-ray bands are presented in Fig.\ 1.
Whereas there is little
indication for extra-planar emission in the hard band, the enhanced emission
in the soft band extends as far as 8 kpc away from the plane of the galaxy,
preferentially towards the north. The diffuse emission becomes 
progressively more extended with decreasing
energy (Fig.\ 2). The distribution of the low surface brightness emission
is nearly symmetric, relative to the minor axis of the galaxy. 
This extra-planar soft X-ray emission clearly represents
the corona of the galaxy.

We have conducted a preliminary X-ray spectral analysis of the corona. 
We extracted a point-source excised spectrum in a corona region 4$^\prime$
wide and 2\farcm5 high (cf. Fig.\ 2b). The spectrum in the 0.3--2 keV range 
can be characterized reasonably well ($\chi^2/{\rm d.o.f.} = 78/47$; Fig.\ 3) 
by a thermal plasma model with two temperature components: 0.18(0.16--0.20) keV
and 0.61(0.48--0.71) keV. The jointly fitted metal abundance and absorption
column density are 8(5--12)\% solar and $3.7(2.2$--$5.8) \times 10^{20}
{\rm~cm^{-2}}$, which is consistent with the expected foreground absorption.
The uncertainties quoted in the parentheses are 90\% confidence intervals.
Assuming that the model is a reasonable characterization of the corona
emission, we derive a conversion ratio of the emission measure (in units
of ${\rm~cm^{-6}~pc}$) to the observed count intensity
($10^{-4} {\rm~counts~s^{-1}~arcmin^{-2}}$) of 91 (27) for the low
(high) temperature component.  The average ratio of the low to high
temperature emission measures is $\sim 8$ over the corona region.
The low-temperature component dominates the outer corona ($\gtrsim 2^\prime$),
whereas the high-temperature component becomes significant in regions close
to the disk. The characteristic gas density and thermal pressure 
($p/k_B$) are
$\sim 1 (5) \times 10^{-3} {\rm~cm^{-3}}$
and $\sim 5 (40) \times 10^3 {\rm~cm^{-3}}$ K in the outer (inner) corona.
The total luminosity of the corona is a few times $10^{39} {\rm~ergs~s^{-1}}$
in the 0.3--2 keV band; the uncertainty in the absorption correction 
is large. More careful analysis is required, especially in regions 
close the disk. Nevertheless, the amount of observed diffuse X-ray emission 
clearly accounts for only a small fraction ($\lesssim 1\%$) of 
the estimated energy input from supernovae in the galaxy (Wang et al. 1995).

The energy-dependent distribution of the diffuse X-ray emission is further
illustrated in Fig.\ 4. The emission in the lowest energy range
(0.3--0.6 keV) has the flattest distribution. The dip just below the
galaxy's major axis ($z = 0'$), most prominent in lower energy bands,
is caused by X-ray absorption from the galactic disk which is 
slightly tilted with its near side below the major axis. The dip at $z = 2'$ 
spatially coincides with a giant dusty arch discovered recently by
Neininger \& Dumke (1999). In fact, this arch seems to enclose
the enhanced soft diffuse X-ray emission, approximately outlined by the outer
white contour in Fig.\ 2a, in the northern central region of the galaxy.

\section{Multiwavelength Comparison}

Fig.~2b presents a morphological comparison of the diffuse 
X-ray and H$\alpha$ emissions from NGC 4631. Whereas the H$\alpha$ line
emission represents primarily warm gas photo-ionized by stars, the soft
diffuse X-ray emission arises in coronal gas. The coronal gas is distributed
further out from the disk than the warm gas, although the galaxy might also
have a very low surface brightness H$\alpha$-emitting halo, which may extend
further out (Donahue, Aldering \& Stocke 1995). The enhanced X-ray emission
arises in regions containing coherent vertical H$\alpha$ filaments, primarily
in the upper central portion of the galactic disk.  The left one quarter of
the disk, though highly disturbed, does not show many vertical H$\alpha$
filaments or much enhanced extra-planar diffuse X-ray emission.

This association between enhanced H$\alpha$/X-ray filamentary features
appears most vividly in Fig.\ 5, which covers the inner $5.5~{\rm kpc}
\times 5.5~{\rm kpc}$ region of the galaxy (cf. Fig.\ 2b). We obtained this deep
H$\alpha$ image using the Wide-Field Planetary Camera 2 (WFPC2) on board the 
{\sl Hubble Space Telescope} (HST): eight orbits narrow-band imaging plus one 
broad-band continuum. The image shows numerous filaments emanating from the 
disk, mostly apparent in the upper-right part of the image. Such coherent 
loop-like features most likely represent ``a froth of merged superbubbles'', 
which were created by massive star clusters, have broken out from the dense
neutral gas of the disk, and are expanding into the halo (e.g., MacLow,
McCray \& Norman 1989; Norman \& Ikeuchi 1989). The overall correlation
between this froth and the enhanced diffuse X-ray emission is apparent.
However, relatively bright loops do not seem to have a configuration
with open tops, as one might expect in galactic chimney models. 
The prominent V-shaped double filaments, just above the large dark cloud near
the center of the image, may not be an opened chimney and most likely
represent walls of many projected loops. X-ray emission within the loops 
is generally enhanced by a factor of up to $\sim 3$.  Furthermore,
the WFPC2 image reveals several bow-shaped filaments far away from
the central plane of the galaxy. The largest and brightest one, as
labeled as ``Arc'' in Fig.\ 5, ``covers'' the top of the V-shaped 
filaments. These H$\alpha$ filaments may represent the illuminated edges of 
in-falling clouds, which could originate in cooled halo gas or in 
instabilities of the dusty arch.

The overall X-ray morphology of the corona (e.g., Fig.\ 1b) 
resembles the well-known radio halo of this galaxy (Hummel \& Dettmar 1990), 
indicating a close link between outflows of hot gas and cosmic ray/magnetic 
field from the galactic disk.  One scenario for the formation of such a
radio halo is that hot gas draws magnetic fields (and cosmic rays) from the
disk into the halo, before being ultimately confined by the magnetic field
tension at large off-plane distances (Wang et al. 1995).  

The unusually low metal abundance inferred from our spectral fit
is apparently an artifact, which could be caused by the poorly calibrated 
instrument response shape below $\sim 1$ keV (\S 2)
and/or by the overly-simplified 
spectral model we have assumed. A realistic temperature distribution
should span a range, which tends to smooth out temperature-sensitive  
spectral features resulted from metal emission lines.  
While our spectral analysis provides an estimate of the temperature range,
a realistic model is yet to be developed for a reliable measurement
of the abundance. More importantly, however, the out-flowing gas may not 
be in an ionization equilibrium state (Breitschwerdt \& Schmutzler 1999). 
We are currently conducting a spatially-resolved spectroscopic analysis 
of the \C\ data and are investigating a self-consistent dynamical
and thermal plasma emission model. 

Now with the corona around NGC 4631 firmly detected, one may naturally ask
questions such as: does the extent of galactic coronae always coincide with
the radio halos?  Is the galaxy-galaxy interaction necessary to generate the
coronae? Do properties of diffuse hot gas depend on the Hubble type of a
galaxy? Ongoing and future observations with the \C\ and \N\ X-ray
observatories will allow us to answer these questions, leading to a better
understanding of our own Galaxy as well as the structure and evolution of
galaxies in general.

\acknowledgments

We thank M. Markevitch and A. Vikhlinin for helping with the 
calibration software and the referee for useful 
comments, especially on existing problems with the spectral response matrices. 
This work is funded by SAO Chandra grant GO0--1150X, STScI grant
GO0--8166.01, and NASA LTSA grant NAG5--8999.

%%=========================================================================

\begin{figure*}[p!]
\unitlength1.0cm
    \begin{picture}(16,8.5) 
\put(0,0){
          \begin{picture}(8.5,8.5)
	\psfig{figure=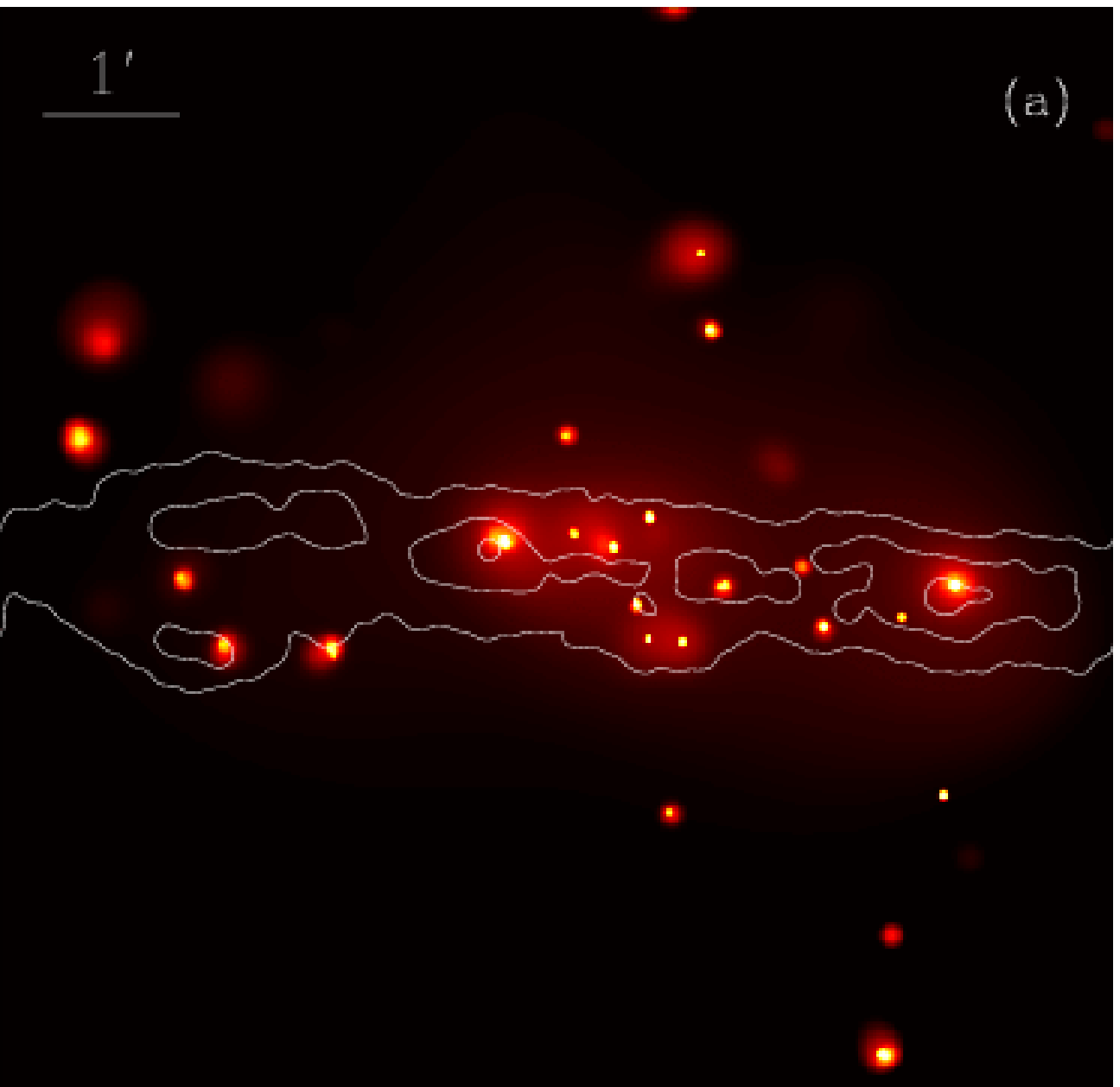,height=8.5cm,clip=}
	\end{picture}
	}
\put(9,0){
          \begin{picture}(8.5,8.5)
	\psfig{figure=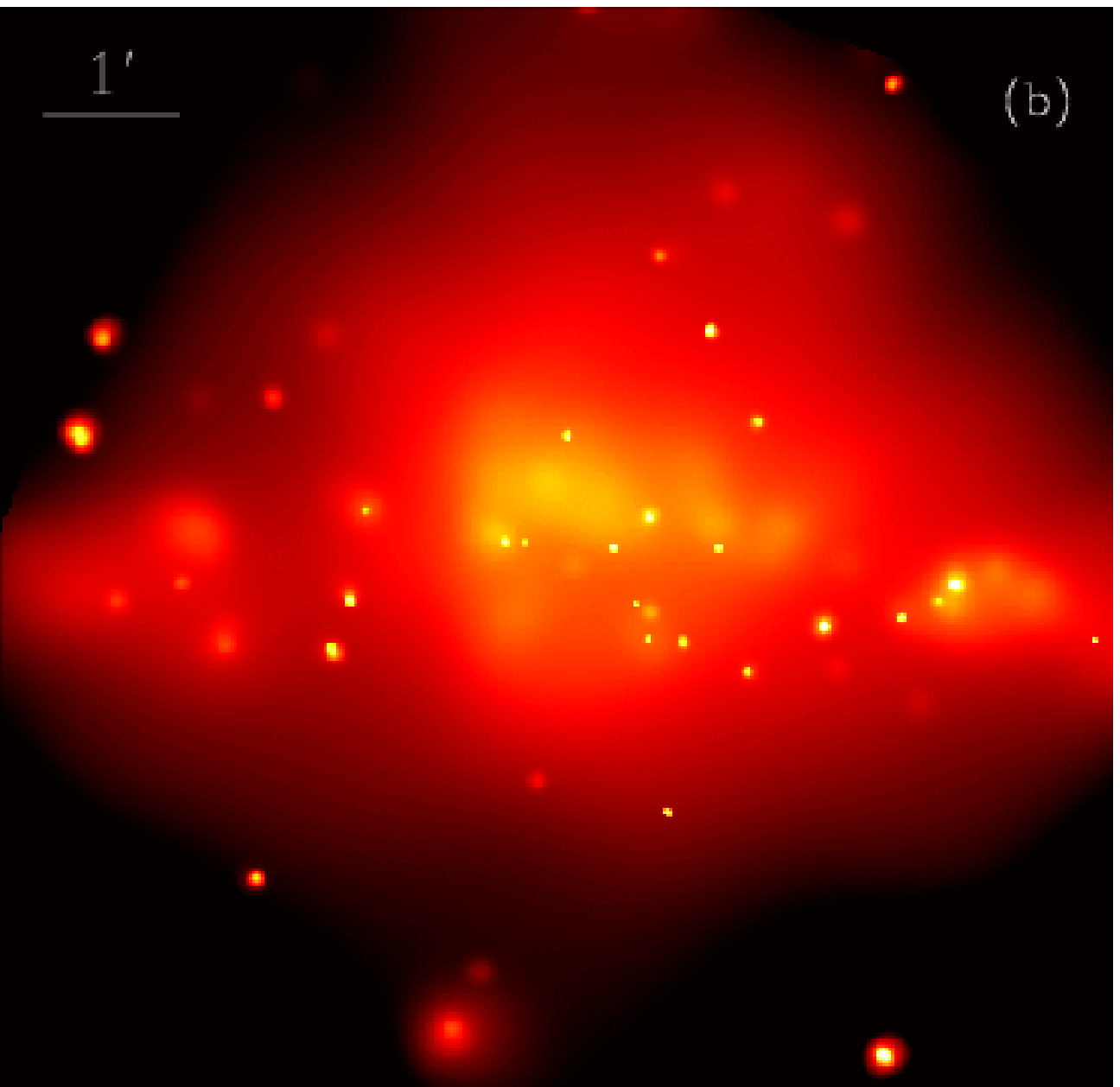,height=8.5cm,clip=}
	\end{picture}
	}
    \end{picture}
\caption{
\C\ ACIS-S images of NGC 4631 in the 1.5--7 keV band (a) and the 0.3--1.5 keV
band (b). These images, both background subtracted and exposure corrected, are 
adaptively smoothed with the program {\sc csmooth} implemented in {\sc ciao}; 
the signal-to-noise ratio of the smoothed image is $\sim3$. The far-UV 
intensity contours outline the morphology of the galactic disk. 
The field-of-view is $18~{\rm kpc} \times18~{\rm kpc}$.
\label{f1}}
\end{figure*}

\figcaption[f2a.ps,f2b.ps]
{Diffuse soft X-ray and H$\alpha$ intensity distributions of NGC 4631. 
The {\sl Chandra} image (a) is produced in the 0.3--0.9 keV
band and discrete sources (marked by {\sl crosses}) are excised.  The
ground-based H$\alpha$ image (b) is reproduced from
Hoopes et al. (1999).    The X-ray contours are at 1.3, 2.5, 4.2, 6.6, 10,
15, 21, 31, $45 \times 10^{-4} {\rm~counts~s^{-1}~arcmin^{-2}}$.
The rectangular box marks the region from which the corona spectral data in 
Fig.\ 3 was extracted, whereas the square box illustrates the location 
of the inner region shown in Fig.\ 5.
\label{f2}}

\begin{figure*}[p!]
\centerline{ {\hfil\hfil
\psfig{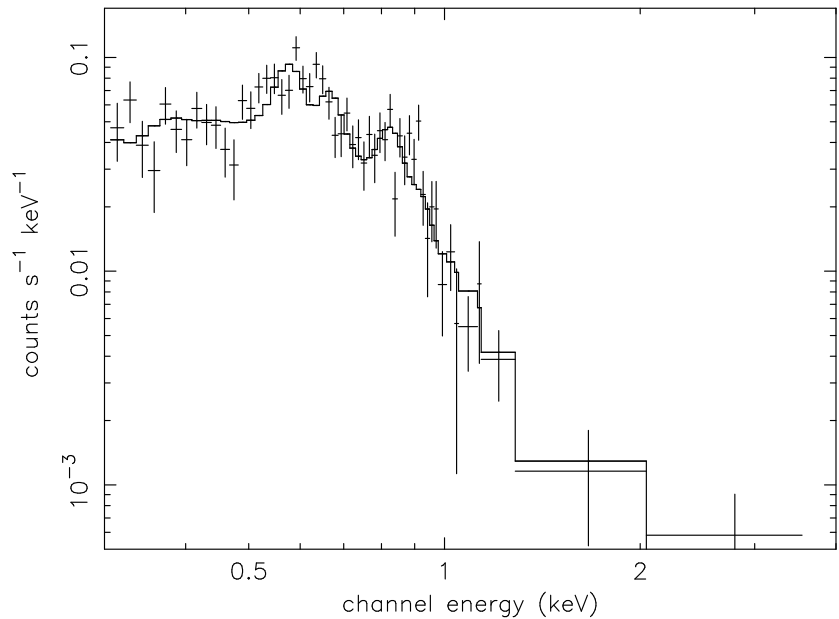}
\hfil\hfil}}
\caption{
\C\ ACIS-S spectrum of the NGC 4631 corona. The histogram represents the 
best-fit two-temperature thermal plasma model (see text).
\label{f3}}
\end{figure*}

\begin{figure*}[p!]
\centerline{ {\hfil\hfil
\psfig{figure=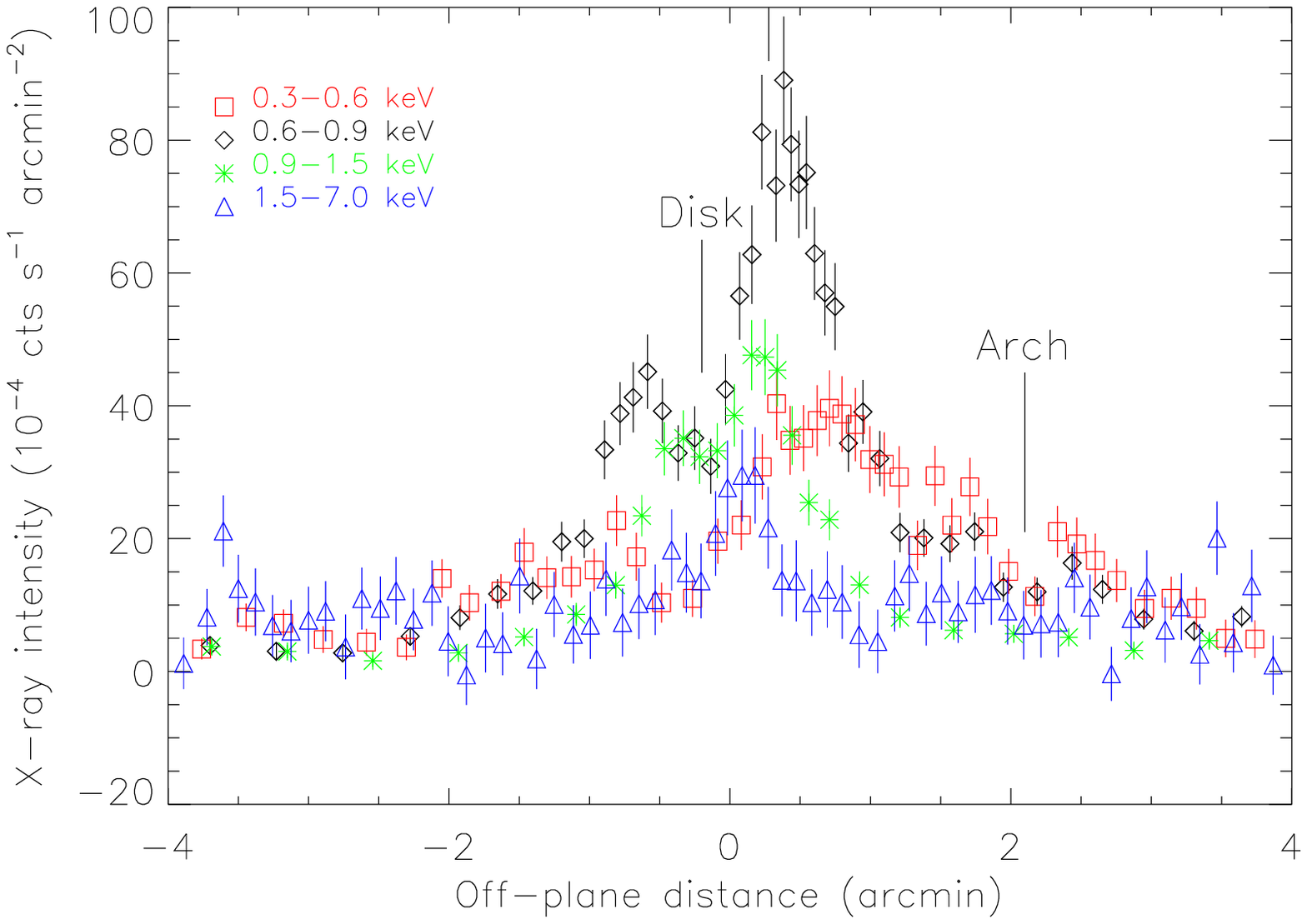,width=11cm,clip=}
\hfil\hfil}}
\caption{
Diffuse X-ray intensity distributions along the minor axis of the galaxy
NGC 4631. The profiles were calculated in a central $2'$ cut perpendicular
to the galaxy's major axis ($86\degr$ eastward of north). The contribution
from discrete X-ray sources marked in Fig.\ 2a was
excised. The northern side of the galaxy is to the right.
\label{f4}}
\end{figure*}

\figcaption[f5a.ps,f5b.ps]
{HST WFPC2 H$\alpha$ image of the central $5.5~{\rm kpc} \times 5.5~{\rm kpc}$
region of NGC 4631 (Fig.\ 2b). High level X-ray contours
in Fig.\ 2 are plotted and a few tentatively identified H$\alpha$-emitting
loops are outlined in right-hand panel. 
\label{f5}}

\end{document}